\documentclass[twocolumn,pre,superscriptaddress,unsortedaddress,showpacs]{revtex4}

\usepackage[intlimits]{amsmath}
\usepackage{amsfonts}
\usepackage{amssymb}
\usepackage{mathtools}

\usepackage{dcolumn}
\usepackage{epsfig}
\usepackage{color}

\usepackage{enumitem}
\usepackage{cleveref}

\newcommand{\integ}{\int \!}
\newcommand{\diff}{\, \mathrm{d}}

\newcommand{\av}[1]{\left<{#1}\right>}

\begin{document}

\begin{abstract}

Exploring the free-energy landscape along reaction coordinates or system parameters $\lambda$ is central to many studies of high-dimensional model systems in physics, 
e.g.\ large molecules or spin glasses.
In simulations this usually requires sampling conformational transitions or phase transitions,
but efficient sampling is often difficult to attain due to the roughness of the energy landscape.
For Boltzmann distributions, crossing rates decrease exponentially with free-energy barrier heights.
Thus, exponential acceleration can be achieved in simulations by applying an artificial bias along $\lambda$ tuned such that a flat target distribution is obtained.
A flat distribution is however an ambiguous concept unless a proper metric is used, and is generally suboptimal.
Here we propose a multidimensional Riemann metric, which takes the local diffusion into account, and redefine uniform sampling such that it is invariant under nonlinear coordinate transformations.
We use the metric in combination with the accelerated weight histogram method, 
a free-energy calculation and sampling method, 
to adaptively optimize sampling toward the target distribution prescribed by the metric.
We demonstrate that for complex problems, such as molecular dynamics simulations of DNA base-pair opening,
sampling uniformly according to the metric, 
which can be calculated without significant computational overhead,
improves sampling efficiency by 50--70\%.
\end{abstract}

\title{Riemann metric approach to optimal sampling of multidimensional free-energy landscapes}

\author{Viveca Lindahl}
\affiliation{Department of Physics and Swedish e-Science Research Center, KTH Royal Institute of Technology, 10691 Stockholm, Sweden}
\author{Jack Lidmar}
\affiliation{Department of Physics and Swedish e-Science Research Center, KTH Royal Institute of Technology, 10691 Stockholm, Sweden}
\author{Berk Hess}
\affiliation{Department of Physics and Swedish e-Science Research Center, KTH Royal Institute of Technology, 10691 Stockholm, Sweden}


\maketitle

\section{Introduction}
Calculating the dependence of the free energy on one or several parameters is often a key step in gaining understanding of a complex system, and many simulation methods have been developed for this purpose.
One difficulty is that the free energy cannot be expressed as a single canonical average, but typically needs to be estimated using several ones.
Additionally, for systems that have complicated free-energy landscapes with high barriers that separate the states of interest, simulations become very challenging due to exponential slowing down when simulated in a canonical ensemble.

A particularly fruitful approach to deal with both these issues in Monte Carlo (MC) or molecular dynamics (MD) simulations consists of extending the ensemble from the ordinary canonical to an \emph{extended ensemble},
where some parameters $\lambda$ are promoted to dynamical variables~\cite{Lyubartsev1992,Marinari1992}.
The parameters in question could be external, like temperature, pressure, or interaction strengths, or more microscopically defined collective variables or reaction coordinates, such as certain interparticle distances, dihedral angles, etc.
The choice of these parameters is often highly specific to the particular problem and rests largely on physical insights or intuition, although it may have a very large impact on the performance of the simulation.
Some attempts to automate the selection of collective variables have been proposed~\cite{McGibbon2017,e2010transition,ma2005automatic,best2005reaction,banushkina2016optimal}, but to a large extent it remains an art rather than a science.
Once selected, an open question is how to distribute the simulated samples (computer resources) among the parameter values $\lambda$ to run the simulation as efficiently as possible.
That is, how to choose the \emph{target distribution} $\pi(\lambda)$
of the extended ensemble.
In sampling methods where multiple independent simulations are performed at different parameter values $\lambda$ or in replica exchange simulations the corresponding problem concerns the placement of the (intermediate) $\lambda$-values~\cite{shenfeld2009minimizing, park2014theory}.
In this paper, we propose that these questions can be approached by considering the geometry induced by an intrinsic Riemann metric defined on the parameter manifold $\Lambda$.

Traditionally, a uniform distribution of samples is targeted~\cite{Lyubartsev1992, Marinari1992, bennett1976efficient,berg1991multicanonical, wang2001efficient, laio2002escaping, steinbrecher2007nonlinear}
irrespective of the chosen parametrization $\lambda$,
i.e.\
$\pi(\lambda) = \text{const.}$ for $\lambda$ in a closed, pre-defined subregion of $\Lambda$.
However, it  has been shown for a variety of applications 
that such choices are suboptimal~\cite{trebst2004optimizing,katzgraber2006feedback,trebst2006optimized, martinez2008variance, buelens2012linear, Tian2014, branduardi2012metadynamics}.
Trebst \emph{et al}.~\cite{trebst2004optimizing} showed,
in the one-dimensional case,
that substantial gains in efficiency may be obtained  by maximizing the probability flow across the parameter range.
The standard way to accomplish this is to keep track of two separate histograms $N_\pm(\lambda)$ of visited parameters for random walks going up or down the parameter range by adding a label $\pm$ to the walker depending on whether the upper ($+$) or lower ($-$) extreme was visited last~\cite{trebst2004optimizing}.
Then a local diffusion constant $D(\lambda)$ is estimated from the probability current with an assumed form
$J = D(\lambda) \pi(\lambda) d q(\lambda)/d\lambda$, where $q(\lambda) = N_+(\lambda)/(N_+(\lambda) + N_-(\lambda))$ is the average fraction of down-walkers.
Optimizing the flow, or equivalently minimizing the mean round trip time~\cite{nadler2007generalized}, gives a distribution 
$\pi(\lambda) \propto 1/\sqrt{D(\lambda)}$.
Although very elegant, this requires the numerical evaluation of a derivative of the global estimate $q(\lambda)$, 
which is reliable only after a considerable amount of sampling and which can be sensitive to the location of the sampling boundaries~\cite{Tian2014}.
Also, it is not obvious how to extend this approach to higher dimensions.
Some generalizations in these directions have been proposed previously,
e.g, local estimates of the diffusion constant for the one-dimensional case~\cite{Tian2014}.
In two dimensions the diffusion constant has been estimated from two marginal one-dimensional histograms~\cite{Singh2012}, which presupposes the absence of correlations and therefore does not generally solve the problem.

Here we propose to optimize $\pi(\lambda)$ 
by endowing $\Lambda$ with an intrinsic Riemann metric,
inducing a geometry dictating which paths are optimal to sample and how to distribute samples along each path.
The metric is a local quantity  and this geometric approach is applicable also for improving sampling in high-dimensional manifolds.
In Sec.~\ref{sec:theory} we derive a suitable choice of  metric by,
rather than maximizing probability flow,
reformulating the optimization problem as one of minimizing the variance of the estimated free-energy difference $\Delta F$ along an arbitrary path connecting two states in $\Lambda$.
In contrast to previous work that combines concepts of information geometry and free-energy calculation~\cite{shenfeld2009minimizing}, 
here we incorporate also dynamic information of the sampling method into the metric.
The optimization can be carried out on the fly, with negligible overhead and without pre-calculating the metric, using an adaptive biasing potential method
as described in Sec.~\ref{sec:methods}.
In Sec.~\ref{sec:applications}
we demonstrate the practical gains of the procedure for the case of calculating the free energy as a function of a reaction coordinate,
the potential of mean force (PMF), in MD simulations.
In all three test cases,
a polymer chain on a surface, ion-pair separation of lithium acetate and DNA base pair opening,
the metric-based optimization shortens the simulation time required to reach the same level 
of statistical accuracy.

\section{Theory}	\label{sec:theory}
\subsection{Extended ensemble simulations}
To begin, we discuss briefly the idea of extended ensemble simulations.
We assume first that we have a simulation method generating (typically correlated) samples $\{x(t)\}$ with an ordinary canonical distribution
$P(x|\lambda) = e^{F_\lambda - E_\lambda(x)}$
at fixed parameter values $\lambda$ \footnote{In order to be as general as possible, we use dimensionless (free) energies.  The dimensionful energies would be obtained by multiplication by the temperature factor $k_BT$.}.
Here $x$ denotes the microscopic state of the system, e.g. the coordinates and momenta $\{\mathbf r_i, \mathbf p_i \}_1^N$ in a MD simulation.
The updates of $x$ at fixed $\lambda$ are then complemented by MC moves that update $\lambda$ at fixed $x$.
By design this results in a stochastic process with a joint equilibrium distribution
\begin{equation}	\label{eq:P(x,lambda)}
P(x,\lambda) = \frac{1}{\mathcal{Z}} e^{f_\lambda - E_\lambda(x)}
\end{equation}
in the extended ensemble,
where $f_\lambda$ are a set of free parameters.
Specifically, we consider parameter moves that select the new $\lambda$ from the conditional distribution~\cite{lidmar2012improving,lindahl2014accelerated}
\begin{equation}		\label{eq:w}
w_\lambda(x) \equiv P(\lambda | x)
= \frac {e^{f_\lambda - E_\lambda(x)}} {\sum_{\lambda'} e^{f_{\lambda'} - E_{\lambda'}(x)}} .
\end{equation}
Alternatively, one may marginalize over $\lambda$ and generate samples from a simulation with an equilibrium distribution $P(x) = \sum_\lambda P(x,\lambda)$ (see also Sec.~\ref{sec:adaptive}).
Integrating out $x$ from the joint distribution
yields 
\begin{equation} \label{eq:plambda}
P(\lambda) = \frac{1}{\mathcal{Z}}e^{f_\lambda - F(\lambda)},
\end{equation}
where  $F(\lambda) = -\ln {\integ \diff x\,e^{- E_\lambda(x) }}$
is the dimensionless free energy at $\lambda$.
Thus, by tuning the  parameters $f_\lambda$ so that $f_\lambda \approx F(\lambda) + \ln \pi(\lambda)$
the marginal distribution will approach the target distribution,
$P(\lambda) 
\approx \pi(\lambda)$.
This is rather nontrivial to accomplish, since the free energy $F(\lambda)$ is usually unknown from the start and needs to be estimated during the course of the simulation.

\subsection{An invariant definition of ``flat''}

Extended ensemble methods are sometimes referred to as flat histogram methods
since  $\pi(\lambda)$ is most commonly chosen uniform,
although, as we have seen above, this may be suboptimal.
In fact, the very notion of a flat distribution is ambiguous unless a metric is specified.
To see this, consider a nonlinear reparametrization  
$\lambda \mapsto \lambda'$,
under which a prescribed target distribution transforms as $\pi(\lambda) d\lambda =\pi(\lambda') d\lambda'$.
This transforms
an originally flat distribution, $\pi(\lambda) = \text{const.}$, into a generally non-uniform one,
$\pi(\lambda') \propto |d\lambda/d\lambda'|$.

We now assume that there is a relevant Riemann metric $g_{\mu\nu}(\lambda)$ defined on the $n$-dimensional manifold of parameters $\Lambda = \{\lambda^\mu\}$.
In practical implementations $\lambda$ is usually discretized, but we assume that the discretization is fine enough that we can use a continuum formulation.
The infinitesimal length and volume elements,
$dl^2 = g_{\mu\nu}(\lambda) d\lambda^\mu d\lambda^\nu$
and
$dV = \sqrt {g(\lambda)} d\lambda^1 d\lambda^2 \ldots d\lambda^n$,
where $g(\lambda) = \det \left( g_{\mu\nu}(\lambda) \right)$ is the determinant of the metric tensor,
are coordinate independent provided that $g_{\mu\nu}$ transforms as a covariant tensor,
\begin{equation}							\label{eq:transf}
g_{\mu\nu}(\lambda) = \frac {\partial \lambda'^\alpha}{\partial \lambda^\mu} \frac {\partial \lambda'^\beta}{\partial \lambda^\nu} g'_{\alpha\beta}(\lambda').
\end{equation}
Here and in the following we use the Einstein summation convention over repeated indices.
The probability measure on $\Lambda$ can then be expressed as
$\pi(\lambda) d\lambda^1 \ldots d\lambda^n = \rho(\lambda) dV$.
This suggests that we  should redefine the notion of ``flat''
to mean $\rho(\lambda) = \text{const.}$
Then
\begin{equation}\label{eq:target}
\pi(\lambda) \propto \sqrt{g(\lambda)}
= \sqrt{\det (g_{\mu\nu}(\lambda))}
\end{equation}
is flat according to the metric $g_{\mu\nu}(\lambda)$.
As we will demonstrate below, such a target distribution can be very useful in extended ensemble simulations.

\subsection{Existing metrics}
Early on, various metrics have been introduced as second derivatives of the thermodynamic potentials $U$ or $S$, and used to study finite time thermodynamic processes~\cite{Weinhold1976,Ruppeiner1979,Salamon1983}.
In the context of probability and statistics a natural measure of distance is the Kullback-Leibler divergence~\cite{kullback1951information}
$D_{\mathrm{KL}}(\lambda \| \lambda') = \int P(x|\lambda) {\ln [P(x|\lambda)/P(x|\lambda')]} dx$.
When expanded to second order one obtains
$D_{\mathrm{KL}}(\lambda + \delta \lambda \| \lambda) \approx \frac 1 2 g^{\text{FR}}_{\mu\nu} \delta\lambda^\mu \delta\lambda^\nu$,
where
\begin{equation}						\label{eq:Fisher-Rao}
g^{\text{FR}}_{\mu\nu}(\lambda) = \int dx P(x|\lambda)
{\partial_\mu \ln P(x|\lambda) \partial_\nu \ln P(x|\lambda)}
\end{equation}
is the Fisher-Rao information metric~\cite{Rao1945},
and $\partial_\mu \equiv \partial / \partial \lambda^\mu$.
The latter has been used to optimize the placement of intermediate states in replica exchange simulations~\cite{shenfeld2009minimizing}.
These metrics however, ignore the time correlations present in the generating process.  Recently, Sivak and Crooks proposed a metric obtained as the leading contribution to the excess work from a slowly externally controlled nonequilibrium process~\cite{sivak2012thermodynamic},
\begin{equation}
						\label{eq:Sivak-Crooks-metric}
g^{\text{SC}}_{\mu\nu}(\lambda) = \int_0^\infty dt  \av{\delta \mathcal F_\mu(x(t),\lambda) \delta \mathcal F_\nu (x(0),\lambda)}_\lambda,
\end{equation}
where the average is taken in equilibrium at fixed $\lambda$ and
\begin{equation}
\delta \mathcal{F}_\mu(x,\lambda) = 
\mathcal{F}_\mu(x,\lambda) - \av{\mathcal{F}_\mu(x,\lambda)}_\lambda =
\partial_\mu \ln P(x|\lambda)
\end{equation}
is the fluctuation of the generalized force
\begin{equation}\label{eq:force} 
\mathcal{F}_\mu(x,\lambda) = - \partial_\mu E_\lambda(x)
\end{equation}
conjugate to $\lambda^\mu$.
The time-integrated force correlation functions
$g^{\text{SC}}_{\mu\nu}$  are the matrix elements of a friction tensor~\cite{sivak2012thermodynamic}.
It generalizes earlier metrics by incorporating time correlations, but does not exactly apply to the situation we are interested in where $\lambda$ changes stochastically rather than
according to an external protocol.

\subsection{Derivation of the metric}	\label{sec:derivation}

Here we derive an intrinsic Riemann metric $g_{\mu\nu}(\lambda)$ on $\Lambda$,
defined in terms of the stochastic process used to generate the samples of the simulation.
We first consider the evaluation of the free-energy difference $\Delta F = F(\lambda_f) - F(\lambda_0)$ along a path $\lambda^\mu(s)$,
\begin{equation}\label{eq:DeltaF}
\Delta F = \int_{\lambda_0}^{\lambda_f} d\lambda^\mu \partial_\mu F(\lambda)
= - \int_{0}^{1} ds \dot \lambda^\mu(s) \av{\mathcal F_\mu(x,\lambda(s))}_{\lambda(s)},
\end{equation}
where the dot denotes a derivative with respect to $s$.
Rather than carrying out simulations that generate samples from $P(x|\lambda)$ at fixed values of $\lambda$,
we consider an extended ensemble $P(x,\lambda)$.
Using samples from an extended ensemble trajectory, $0 \le t \le \tau$,  we may estimate
$\Delta F$ as
\begin{equation}						\label{eq:dF-estimate}
\overline{\Delta F}
= - \int_0^1 ds \frac {\dot \lambda^\mu(s)} {\tau P(\lambda(s))}
\int_0^\tau dt \mathcal F_\mu(x(t),\lambda(s)) w_{\lambda(s)}(x(t)),
\end{equation}
where $w_\lambda(x)$, see Eq.~\eqref{eq:w}, reweights the samples $x(t)$ to the $\lambda^\mu(s)$ of interest.
It is easy to see that $\overline{\Delta F}$ is an unbiased estimator of $\Delta F$, $\av{\overline{\Delta F}} = \Delta F$.
Its variance becomes
\begin{widetext}
\begin{equation}					\label{eq:Var}
\mathrm{Var}\ \overline{\Delta F} = \av{(\overline{\Delta F} - \Delta F)^2}
= \int_0^1 ds' \int_0^1 ds
\frac{\dot \lambda'^\mu \dot \lambda^\nu}{\tau^2 P(\lambda') P(\lambda)}
\int_0^\tau dt' \int_0^\tau dt
\av{\delta \mathcal{F}_\mu(x',\lambda') \delta \mathcal{F}_\nu(x,\lambda) w_{\lambda'}(x')w_{\lambda}(x)},
\end{equation}
\end{widetext}
where $\lambda' \equiv \lambda(s')$, $\lambda \equiv \lambda(s)$,
and the average is taken with respect to $P(x',t' ; x,t)$,
the joint two-time equilibrium distribution of the extended ensemble simulation.
The resulting expression is clearly quite complicated and nonlocal,
linking spatial and temporal correlations.
In many cases of interest, however, we expect the integrand to be sharply peaked so that the main contribution will come from the diagonal elements $s' \approx s$.
This rests on the assumption that the overlap of $w_{\lambda'}(x(t))$ and $w_\lambda(x(0))$ is negligible unless $\lambda' \approx \lambda$
and that the generalized forces tend to decorrelate when going from  $\lambda$ to a distant $\lambda'$.
For parameters $\lambda$ representing macroscopic properties  or when the dynamics of $\lambda$ is slow compared to the microscopic degrees of freedom,
properties which are commonly perceived as desirable for reaction coordinates,
this should often be a  reasonable approximation.

Here assuming this to be the case, we will now derive an approximate coarse-grained expression, local in $s$, by extracting the dominating factor
$P(\lambda',\epsilon ; \lambda,0) = \av{w_{\lambda'}(x(\epsilon))w_{\lambda}(x(0))}$
from the integrals, where $\epsilon$ is a short time scale such that this probability is highly peaked around $\lambda' \approx \lambda$, but large enough that it is positive for all $\lambda'$.
We multiply and divide by this factor and in addition
take the long time limit $\tau \gg 1$. Using that the average is stationary,
the double time integral reduces by symmetry to  a one-dimensional integral over lag times and a factor of $2\tau$,
\begin{widetext}
\iffalse % alternative, single equation
\begin{equation}						\label{eq:Var2}
\mathrm{Var}\ \overline{\Delta F} =
\frac{2}{\tau} \int_0^1 \! ds' \int_0^1 \! ds
\dot \lambda'^\mu \dot \lambda^\nu
\frac{P(s',\epsilon | s,0)}{P(s')}
\int_0^\infty \! dt
\frac
{\av{\delta \mathcal{F}_\mu(x(t),\lambda') \delta \mathcal{F}_\nu(x(0),\lambda) w_{\lambda'}(x(t))w_{\lambda}(x(0))}}
{P(\lambda', \epsilon ; \lambda,0)} .
\end{equation}
\fi
\begin{align}						\label{eq:Var2}
\mathrm{Var}\ \overline{\Delta F} &=
\frac{2}{\tau} \int_0^1 \! ds' \int_0^1 \! ds
\frac{\dot \lambda'^\mu \dot \lambda^\nu P(\lambda', \epsilon ; \lambda,0)}{P(\lambda') P(\lambda)}
\int_0^\infty \! dt
\frac{
\av{\delta \mathcal{F}_\mu(x(t),\lambda') \delta \mathcal{F}_\nu(x(0),\lambda) w_{\lambda'}(x(t))w_{\lambda}(x(0))}
}
{P(\lambda', \epsilon ; \lambda,0)}
\nonumber
 \\
&=
\frac{2}{\tau} \int_0^1 \! ds' \int_0^1 \! ds
\dot \lambda'^\mu \dot \lambda^\nu
\frac{P(s',\epsilon | s,0)}{P(s')}
\int_0^\infty \! dt
\frac
{\av{\delta \mathcal{F}_\mu(x(t),\lambda') \delta \mathcal{F}_\nu(x(0),\lambda) w_{\lambda'}(x(t))w_{\lambda}(x(0))}}
{P(\lambda', \epsilon ; \lambda,0)},
\end{align}
\end{widetext}
where in the final step we have used
$P(\lambda', \epsilon ; \lambda,0)/P(\lambda')P(\lambda) =P(s', \epsilon | s,0)/P(s')$ 
(from the definition of conditional probabilities and changing variables from $\lambda$ to $s$).
Assuming that the rest of the integrand varies slowly with $s'$ compared to
the sharply peaked
$P(s', \epsilon | s,0) \approx \delta(s'-s)$,
we may approximate the integral over $s'$ with the result
\begin{equation}						\label{eq:finalVar}
\mathrm{Var}\ \overline{\Delta F} \approx 
\frac{2}{\tau} \int_0^1 ds \frac {\dot \lambda^\mu(s) \dot \lambda^\nu(s) g_{\mu\nu}(\lambda(s);\epsilon)}
{P(s)} ,
\end{equation}
where
\begin{equation}							\label{eq:metric1}
g_{\mu\nu}(\lambda;\epsilon) =
\int_0^\infty dt
\frac{
\av{\delta \mathcal{F}_\mu(x(t),\lambda)
\delta \mathcal{F}_\nu(x(0),\lambda)
w_\lambda(x(t)) w_\lambda(x(0))}}
{P(\lambda,\epsilon ; \lambda,0)}
.
\end{equation}
Note that the $1/\tau$ decay of the variance found here holds generally, irrespective of the target distribution and the density of $\lambda$-values (as long as the spacing is small enough), in conformance with previous findings~\cite{nguyen2016intermediate}.
In the derivation so far we have allowed the weight factors $w_\lambda(x)$ to be general.
Instead of reweighting one could, e.g., use a simple histogram estimator by 
replacing all occurrences of $w_\lambda(x(t))$ with $w_{\lambda,\lambda(t)} = \delta(\lambda - \lambda(t))$
in Eqs.~\eqref{eq:dF-estimate}-\eqref{eq:metric1}.
Using reweighting, however, the sampled data is used more efficiently.
Often it is practical to work with weights that have support on the whole simulated parameter range, i.e., $w_{\lambda}(x) > 0 \ \forall \lambda(s)$, 
so that $P(s', \epsilon | s, 0)$ will remain finite and positive for all $\epsilon > 0$,
and we assume this to be the case from now on
(this certainly holds for the choice in Eq.~\eqref{eq:w}).
Then there is no lower bound for $\epsilon$,
and we may safely take the limit $\epsilon \to 0^+$,
i.e.\ 
$
P(\lambda, \epsilon \to 0^+; \lambda, 0) = \av{w_\lambda^2(x)},
$
so that we finally arrive at the expression we will use as our \emph{metric} on the parameter manifold
\begin{equation}							\label{eq:metric}
g_{\mu\nu}(\lambda) =
\int_0^\infty 
dt
\frac{
\av{\delta \mathcal{F}_\mu(x(t),\lambda)
\delta \mathcal{F}_\nu(x(0),\lambda)
w_\lambda(x(t)) w_\lambda(x(0))}}
{\av{w^2_\lambda(x)}}
.
\end{equation}
The symmetric tensor $g_{\mu\nu}$ is positive definite and so indeed defines a metric~\footnote{Zero eigenvalues may occur if the parametrization is redundant.  We assume this not to be the case.}.
Furthermore, it is readily verified that it satisfies Eq.~\eqref{eq:transf} and therefore takes the same form under arbitrary parameterizations~\footnote{Note that $g_{\mu\nu}$ defines a metric regardless of the validity of the approximations performed.  In fact, Eq.~\eqref{eq:metric1} defines a family of metrics parametrized by $\epsilon$.}.

\subsection{Optimal sampling with the metric}\label{sec:optimize}
By using the Cauchy-Schwarz inequality, 
in the form $\int f^2(s) /\rho(s) ds \int \rho(s) ds \geq \left(\int f(s) ds \right)^2$,
we may bound the approximate variance in Eq.~\eqref{eq:finalVar} as
\begin{equation}					\label{eq:Cauchy-Schwarz}
\mathrm{Var}\ \overline{\Delta F} \geq  2 \frac {\mathcal L^2} \tau ,
\end{equation}
where
\begin{equation}					\label{eq:Length}
\mathcal L = \int_0^1 ds
\sqrt{g_{\mu\nu}(\lambda(s)) \dot \lambda^\mu(s) \dot \lambda^\nu(s)}
\end{equation}
is the length of the curve.
The equality occurs for $P(s) \propto \sqrt{g_{\mu\nu}(\lambda(s)) \dot \lambda^\mu(s) \dot \lambda^\nu(s)}$,
which is the optimal target distribution for a fixed one-dimensional path $\lambda^\mu(s)$.
We assume here that the computational cost is the same for all $\lambda$~\footnote{With a cost $c(s)$ the optimal distribution instead becomes $P(s)/\sqrt{c(s)}$~\cite{trebst2004optimizing}.}.
Obviously, the path giving the lowest error in the free-energy estimate is the geodesic, i.e., the path with the shortest length connecting the states of interest.
The geodesics allow us to define an intrinsic distance
\begin{equation}			\label{eq:distance}
d(\lambda_1, \lambda_0) = \inf_{\lambda(s)}
\left\{ \, \mathcal L[\lambda] \, : \;
\lambda(1) = \lambda_1 ,\, \lambda(0) = \lambda_0 \, \right\}.
\end{equation}Finding the geodesic path is, however, nontrivial, unless we know the metric on the whole parameter space beforehand.
In many cases we instead have to be content with freely exploring a relatively low-dimensional space $\Lambda$.
In doing so, we propose that sampling according to a ``flat'' target distribution,
as defined by Eq.~\eqref{eq:target} using the metric of Eq.~\eqref{eq:metric},
is beneficial.

For the simplest one-dimensional case,
where $\Lambda$ is restricted to an interval of length
$\Delta \lambda =|\lambda_f-\lambda_0|$,
we can estimate the expected improvement from optimizing the target distribution as follows.
Writing Eq.~\eqref{eq:finalVar} using $s = \lambda$ as the parametrization,
we obtain the variance (after also taking the limit $\epsilon\to 0$)
\begin{equation}\label{eq:sigma2}
\sigma^2 \equiv \mathrm{Var}\ \overline{\Delta F}_{\mathrm{1D}} \approx
\frac{2}{\tau}
\int_{\lambda_0}^{\lambda_f} \! \diff \lambda \, \frac{g(\lambda)}{P(\lambda)}.
\end{equation}
Thus, in the unoptimized flat case where $P(\lambda) = 1/\Delta\lambda$,
we obtain  $\sigma^2_{0} = c \overline{g}$, where $c=2\Delta\lambda^2/\tau$ and 
the bar denotes an arithmetic average over $\Lambda$.
In the optimized case, $P(\lambda)\propto \sqrt{g(\lambda)}$, we obtain
$\sigma^2_{\mathrm{opt}}=c \left(\overline{\sqrt{g}}\right)^2$.
So we estimate the optimization to  reduce  the variance by a factor of
\begin{equation}\label{eq:improvement}
\frac{\sigma^2_{\mathrm{opt}}}{\sigma^2_{0}}=
\frac{\left(\overline{\sqrt{g}}\right)^2}{\overline{g}}.
\end{equation}
In a real application, 
the improvement may be higher or lower due to the approximations that have been made in defining the metric
and the sampling method.

In the multidimensional case it is less clear what to consider optimal.
The ``flat'' target distribution \eqref{eq:target} is arguably a good choice if all points of $\Lambda$ in the free-energy landscape are of equal interest, but this is seldom the case.
Often certain regions may correspond to conflicting parameters and unphysical situations with correspondingly large free energies.
Also, the computational cost may become prohibitively high as the number of dimensions is increased.
One way to remedy this is to restrict the sampling to regions with relatively low free energy, counted e.g., from the global minimum~\cite{lindahl2014accelerated}.
The introduction of a free-energy dependent cutoff in the target distribution is readily combined with the metric, giving a target distribution of the form
\begin{equation}			\label{eq:target-cutoff}
	\pi(\lambda) \propto \sqrt{g(\lambda)}\, \varphi(F(\lambda)-F_{th}),
\end{equation}
with e.g., $\varphi(z) = \min \left\{ 1, \exp(-z) \right\}$ or a smoother
$\varphi(z) = 1/(1 + \exp(z))$, and where $F_{th}$ is a stipulated threshold free energy,
below which sampling should be uniform.

Another idea is to use the metric distance \eqref{eq:distance} to limit the sampling to the vicinity of one or more points of interest, e.g., by setting
\begin{equation}			\label{eq:target-distance-1}
\pi(\lambda) \propto \sqrt{g(\lambda)} \, \exp( - d(\lambda, \lambda_0)/l_0)
\end{equation}
or
\begin{equation}			\label{eq:target-distance-2}
	\pi(\lambda) \propto \sqrt{g(\lambda)} \,
	\exp( - [d(\lambda , \lambda_0) + d(\lambda , \lambda_1)]/l_{01}),
\end{equation}
etc., for some suitable values of $\lambda_{0,1}$, $l_0$, $l_{01}$.
The contours of the latter choice are (hyper) ellipsoids with focal points at $\lambda_{0,1}$ deformed by the metric.
The resulting tube shaped region may be useful to map out reaction pathways.

\subsection{Properties of the metric}\label{sec:metric-notes}

A few  additional points are worth noting:

(i) The metric is given by an integrated time-correlation function, and therefore proportional to the correlation times present in the problem. Regions with slow dynamics will thus have a large metric. This can help identifying problematic transitions. If we choose a target distribution as in Eq.~\eqref{eq:target}, more samples will automatically be allocated in these regions.

(ii)
The local approximation made in going from Eq.~\eqref{eq:Var} to \eqref{eq:finalVar} is essentially a Markov approximation.
While not always justified, a good choice of parameters often consists of slow degrees of freedom, where a time-scale separation naturally leads to an effective Markov dynamics.
The approximation is compatible with the one used by Trebst \emph{et al.}~\cite{trebst2004optimizing}, who assume a local relation between probability current and diffusion constant.
Indeed, the metric tensor can be interpreted as the inverse of the diffusion tensor.
In the one-dimensional case we then recover the flow-optimized target distribution 
$\pi(\lambda)\propto1/\sqrt{D(\lambda)}$ of Trebst \emph{et al.},
although the methods estimate the diffusion constant in completely different ways.

(iii) If instead of considering an extended ensemble where $\lambda$ carries out a random walk, we consider a simulation in which $\lambda$ slowly changes deterministically along $s$, and repeat the derivation above for that case, 
we obtain the metric $g^{\text{SC}}_{\mu\nu}(\lambda)$ of Sivak and Crooks, Eq.~\eqref{eq:Sivak-Crooks-metric}.
The same is true for a sequence of equilibrium simulations with different fixed parameters, e.g., in constrained simulations or umbrella sampling.
Assuming proper equilibration has taken place at each $\lambda$, the derivation of Eqs.~\eqref{eq:finalVar} and \eqref{eq:sigma2} using $g_{\mu\nu}^{SC}(\lambda)$ becomes exact in that case.
Like $g^{\text{SC}}_{\mu\nu}(\lambda)$, our metric involves a time-correlation function of the generalized force.
A difference is that the former is evaluated in an ensemble with fixed $\lambda$.
In the extended ensemble the $\lambda$-fluctuations typically help the system equilibrate and may to some extent reduce the correlation time~\cite{chodera2011replica}.  Hence, the metric and thereby the optimized target distribution Eq.~\eqref{eq:target} can be expected to be smoother, reflecting the
advantages of extended ensemble simulations compared to a sequence of fixed-$\lambda$ simulations.

(iv)
The metric  is a function of the dynamics of the simulation algorithm and thereby itself depends on the target distribution.
This means that the optimization of the target distribution should be performed self-consistently.
Fortunately, this may be quite straightforwardly implemented within an adaptively optimizing sampling framework as will be described next.

\section{Methods}\label{sec:methods}
\subsection{Adaptively optimized sampling}
\label{sec:adaptive}
Optimizing $\pi(\lambda)$ adaptively as $g(\lambda)$ is being estimated during the simulation
(according to a procedure described in Sec~\ref{sec:numeric}),
requires choosing a suitable framework. 
Here we use the accelerated weight histogram method (AWH) \cite{lidmar2012improving,lindahl2014accelerated},
an extended ensemble method that updates the required weight functions
$f_\lambda=f_\lambda(t)$
(see Eq.~\eqref{eq:P(x,lambda)}) on the fly
such that sampling along $\lambda$ converges to the chosen target distribution,
$P(\lambda)\to \pi(\lambda)$.
The ensemble is thus time-dependent, but in the following we will, for ease of notation,  leave this dependence implicit when possible.

We consider now the special case when transitions along a reaction coordinate $\xi(x)$ is of interest,
for the  sake of concreteness and since we present numerical results for such applications in Sec.~\ref{sec:applications}.
An extended ensemble may then be defined by coupling each dimension $\xi^\mu(x)$
to a harmonic potential with center at $\lambda^\mu$ and force constant $k_\mu$,
\begin{equation}
E_\lambda(x) = E(x) + \frac{1}{2}  \sum_\mu k_\mu(\xi^\mu(x)-\lambda^\mu)^2,
\label{eq:extendedpot}
\end{equation}
where $E(x) = \beta V(x)$ is the unbiased potential energy divided by temperature $k_B T = 1/\beta$,
and $\lambda^\mu$ takes discrete values on a fine grid.
Thus, here the target distribution is obtained  by adjusting the weights $f_\lambda$ while keeping a fixed grid spacing,
which has the advantage that data collected at different target distributions may be straightforwardly combined.
Alternatively, one could achieve a similar result by instead adapting the grid spacing.
The generalized force, see Eq.~\eqref{eq:force}, corresponding to Eq.~\eqref{eq:extendedpot} simplifies to
\begin{equation}\label{eq:force-reactioncoord}
\mathcal F_\mu(x,\lambda) = k_\mu (\xi^\mu(x) - \lambda^\mu)
\qquad \text{(no sum)}.
\end{equation}
The free energy as a function of $\lambda$, $F(\lambda)=-\ln {\integ \diff x\,e^{- E_\lambda(x)}}$ is
calculated during the AWH simulation.
This is a convolved version of
the PMF along $\xi$,
$\Phi(\xi) = -\ln {\integ \diff x\, e^{-E(x)} \delta(\xi-\xi(x)) }$,
which is  simultaneously extracted by AWH~\cite{lindahl2014accelerated}.

For large force constant $k$ the harmonic potential will restrain $\xi \approx \lambda$ and
$F(\lambda) \approx \Phi(\xi)$.
The metric \eqref{eq:metric} will remain finite in the limit $k_\mu \to \infty$, in contrast to the Fisher-Rao metric Eq.~\eqref{eq:Fisher-Rao}, which using Eq.~\eqref{eq:force-reactioncoord} takes the form
$g_{\mu\nu}(\lambda) = k_\mu \delta_{\mu\nu} - \partial_\mu \partial_\nu F(\lambda)$
and thus is dominated by the trivially flat first term for large $k_\mu$.

In  previous MD work using AWH \cite{lindahl2014accelerated, lindahl2017sequence},
$x$ was sampled at the current $\lambda(t)$,
which in turn was regularly updated using Gibbs sampling,
i.e.\ drawn from
\begin{equation}					\label{eq:w-reactioncoord}
w_\lambda(x) \equiv P(\lambda | x)
= \frac{e^{f_\lambda - \frac 1 2 \sum_\mu k_\mu (\xi^\mu(x) - \lambda^\mu)^2}}{\sum_{\lambda'} e^{f_{\lambda'} - \frac 1 2 \sum_\mu k_\mu (\xi^\mu(x) - \lambda'^\mu)^2}},
\end{equation}
see Eq.~\eqref{eq:w}.
As an alternative, we here instead
sample $x$ from its marginal distribution,
$P(x)
= {e}^{- E(x) - V_b(\xi(x))}/\mathcal{Z}$,
where
$V_b(\xi) = - \ln \sum_{\lambda} {e}^{f_\lambda - \frac{1}{2} \sum_\mu k_\mu (\xi^\mu(x)-\lambda^\mu)^2}$
is the  bias potential  consistent with the current  $f_\lambda$,
which avoids possible high-frequency issues due to choosing large force constants $k_\mu$.
Samples of $\lambda$ may then be drawn when needed from $w_\lambda(x)$.
Thus, in this formulation the extended ensemble is a framework for the inner machinery of AWH, which the MD simulation experiences only through
the time-dependent bias potential  $V_b(\xi)$.

AWH keeps an estimate  of the free energy $\hat{F}(\lambda)$
that is regularly updated using  samples $w_\lambda(x(t))$ collected in between updates.
After updating $\hat{F}(\lambda)$,  the target $\pi(\lambda)$ may be optimized, 
here according to Eq.~\eqref{eq:target} or \eqref{eq:target-cutoff} given an estimate of the metric.
Finally, the bias function $f_\lambda$ is tuned consistently with Eq.~\eqref{eq:plambda}, $f_\lambda = \ln \pi(\lambda) + \hat{F}(\lambda)$,
after which sampling proceeds in the updated ensemble.
Explicitly, with $n$ samples $x(t_i)$ taken at times $t_i$ since the last update, 
the free-energy update is given by
\begin{equation*}
\hat{F}_{\text{new}}(\lambda)=
 \hat{F}_{\text{old}}(\lambda)
-\ln{
\frac
{W_{\text{ref}}(\lambda) + \sum_{t_i}{w}_\lambda(x(t_i)) }
{W_{\text{ref}}(\lambda) + n\pi(\lambda) }.
}
\end{equation*}
$W_{\text{ref}}(\lambda) = \sum_{t' < t} \alpha(t')\pi(\lambda,t')$,
is a reference weight histogram  representing the whole targeted sampling history.
Its normalization determines the overall magnitude of the free-energy update.
The scaling factor $\alpha(t)$ sets the effective weight for samples collected at time $t$.
For sake of robustness, we use an initial ``burn-in'' stage~\cite{gromacsmanual}
where the growth of $W_{\text{ref}}$ is artificially restricted
and consistently, the weights of early samples are scaled down, i.e.\ $\alpha < 1$.
After exiting the initial stage, all samples are weighted equally, $\alpha =  1$.
Employing this type of two-stage algorithm can be critical for attaining efficient convergence~\cite{henin2004overcoming, fort2015convergence, tan2017optimally}.
In the final stage, as $W_{\text{ref}}$ grows linearly with time,
the magnitude of the free-energy update decreases  as $\sim 1/t$
and $\hat{F}$ is expected to converge as $1/\sqrt{t}$~\cite{lindahl2014accelerated}.
Ordinary canonical averages may be calculated during the simulation by taking into account the time-dependent sample weights and
removing the time dependent bias $V_b(\xi,t)$,
\begin{equation}			\label{eq:reweighting}
\bar{A}_\text{can} = 
\frac{\sum_t A(x(t)) \alpha(t) e^{V_b(\xi(x(t)),t)}}
{\sum_t \alpha(t) e^{V_b(\xi(x(t)),t)}}.
\end{equation}

\subsection{Numerical calculation of the metric}
\label{sec:numeric}
Given  samples $\{x(t)\}_{t\in S}$ taken at times $S=\{0,\ldots,T\}$ from a trajectory of length $T$,
we may estimate the metric
using time averages.
Assuming stationarity of the average and time-reversibility
 we have for a  time-correlation  function $C(t,t')=\av{\delta X(t)\delta Y(t')}$,
\begin{align}
  \int_0^{T} \! \diff t \, \delta X(t) \int_0^{T} \! \diff t' \, \delta Y(t') 
&\approx
2 T
\int_0^{T} \! \diff t \, C(t,0)\gamma_{\Delta}(t)\\
&\approx
2T\int_0^{T} \! \diff t \, C(t,0),
\end{align}
where $\gamma_{\Delta}(t) = |1-t/T|$, for $0<t<T$,  is a triangular window function of half width $T$,
and the last approximation is valid for large $T$ such that $C(t,0)$ has decayed sufficiently on a time scale $\lesssim T$.
Applying this to our case, Eq.~\eqref{eq:metric},
together with
$\av{w^2_\lambda(x)}\approx \sum_{t\in S} w^2_\lambda(x(t))/T$,
we thus obtain an estimate of the metric,
\begin{equation}
\hat{g}^S_{\mu\nu}(\lambda) =
\frac{\Delta t}{2}
\frac{I_\mu^S(\lambda)
I_\nu^S(\lambda)
}{I_2^S(\lambda)},
\label{eq:metricestimateblock}
\end{equation}
where we have defined the sums
$I_\mu^S(\lambda) =
\sum_{t\in S}
\delta \mathcal{F}_\mu(x(t),\lambda)w_\lambda(x(t))$
and
$I_2^S(\lambda)=
\sum_{t\in S} w^2_\lambda(x(t))$,
and $\Delta t$, the sampling time interval,
comes from discretizing the integrals.

We expect our simulations to be significantly longer than the correlation times we are interested in sampling.
Thus, to get a consistent and more robust estimate,
we partition the full trajectory
into $N_b$ disjoint blocks $S_i$ of equal length in time $T$
and calculate an estimate of the metric  for each block
using  Eq.~\eqref{eq:metricestimateblock} with $S=S_i$.
Our final estimate of the metric is obtained as an average over the blocks,
weighting each block $S_i$ by $I_2^{S_i}$,
\begin{equation}
  \label{eq:blockaverage}
\hat{g}_{\mu\nu}(\lambda) =
\frac{\Delta t}{2}
\frac{1}{I_2(\lambda)}
\sum_{i = 1}^{N_b}
I_\mu^{S_i}(\lambda)
I_\nu^{S_i}(\lambda),
\end{equation}
where $I_2(\lambda) = I_2^{\cup_i S_i}(\lambda)$
is the sum of squared weights including samples in  all blocks.

Here we determine the  block length $T$ adaptively by
doubling $T$ when 64 blocks have been filled,
i.e.\ when $t > 64T$.
In practice this means that the metric is computed using
$33 \le N_b \le 64$,
depending on how long time has passed since the last doubling.

As was noted in Sec.~\ref{sec:metric-notes},
when optimizing the target distribution $\pi(\lambda)$ with $\hat{g}_{\mu\nu}(\lambda)$ according to  Eq.~\eqref{eq:target},
we are modifying the ensemble and the time correlations present in our samples,
thus also changing the metric itself.
In actual applications one might often have no or little prior knowledge of
the metric. In such cases one would like to estimate the metric and use
it to optimize the target distribution on the fly.
Thus, the question naturally arises
when to  update $\pi(\lambda)$ and 
how to combine samples from different $\pi(\lambda)$ in the metric estimate.
The simplest way is to 
use all data in Eq.~\eqref{eq:blockaverage}
and continuously, at regular intervals, update $\pi(\lambda)$. 
Another, generally more stable way, is to only apply Eq.~\eqref{eq:blockaverage} for samples at a constant  $\pi(\lambda)$,
such that we obtain a set of estimates $\{\hat{g}^n_{\mu\nu}\}$ from samples collected at different targets $\{\pi^n_{\lambda}\}$.
For this scheme,
we update the target at times  when the block length  doubled and then start calculating a new metric estimate for the new target.
This way, the time sampled at fixed $\pi^n(\lambda)$ is proportional $2^n$.
The different metric estimates are combined as
a weighted average with weights $I^n_2(\lambda)$.
Thus, the weight of early metric estimates will rapidly become negligible.
To summarize, we consider the three different optimization  protocols:
\textit{static}:  $\hat{g}$ is pre-calculated from unoptimized AWH simulations, as an average over the simulations, and $\pi(\lambda)$ is constant throughout the simulation;
\textit{dynamic, continuous}: $\hat{g}$ is calculated on the fly using all data and continuously used to update $\pi(\lambda)$;
or \textit{dynamic, doubling}:  $\pi(\lambda)$ is updated every time the block length used in the calculation of the metric is doubled
and $\hat{g}$ is given by a weighted average of a sequence of $\{\hat{g}^n\}$, each calculated at constant $\pi^n(\lambda)$.

\section{Applications}\label{sec:applications}
We now test the use of the metric $g_{\mu\nu}(\lambda)$ for setting the target distribution $\pi(\lambda)$ and improving sampling
for three atomistic systems sampled using MD simulations.
Both the AWH method and the metric calculation for reaction coordinates were implemented in
the molecular simulation software GROMACS~\cite{abraham2015gromacs}
and has been made available in the  2018 release.
For these test systems we calculate the PMF $\Phi(\xi)$
as a function of one or two
reaction coordinates $\xi(x)$.
In this case, the generalized force is given by Eq.~\eqref{eq:force-reactioncoord}.
To obtain good efficiency for the AWH method, $k_\mu$ has to be chosen larger
than the curvature of the PMF.
In addition, as we will see in the applications, the metric often shows sharper features
than the free energy.
In order to obtain maximal improvement in sampling, $k_\mu$ should be chosen large enough to fully resolve also the metric.
In AWH there is no significant computational overhead to increasing $k_\mu$ and
the number of grid points, so one can choose $k_\mu$ as large as the time step
chosen for the integration of the system allows.
Here we use a grid spacing of $1/\sqrt{k_\mu}$.
This grid spacing provides sufficient overlap between  neighboring $\lambda$ points.
Using a finer spacing improves the resolution
along the reaction coordinate, but does not affect the accuracy.

Having decided an application, we need to decide on an error measure for the free energy.
One approach is to use the spatial average of the error in $\Lambda$.
This immediately brings up the question of what metric to use
for the averaging. We would argue this should be the metric we propose here.
But in practice one usually applies enhanced sampling to study transitions
between two (or more) states. For the one dimensional case this naturally leads to
using the root mean square (RMS) error of the free-energy difference between the extreme
values of the reaction coordinate. This is also exactly what using the metric
as target distribution optimizes for. This approach does not immediately generalize
to higher dimensions. So for our two-dimensional application we measure the RMS error of
the free-energy difference between two local minima (which are located close to the
extremes of the reaction coordinates).

\subsection{Example I: A polymer chain on a surface}\label{sec:polymer}
\begin{figure}
\includegraphics[height=5.5cm]{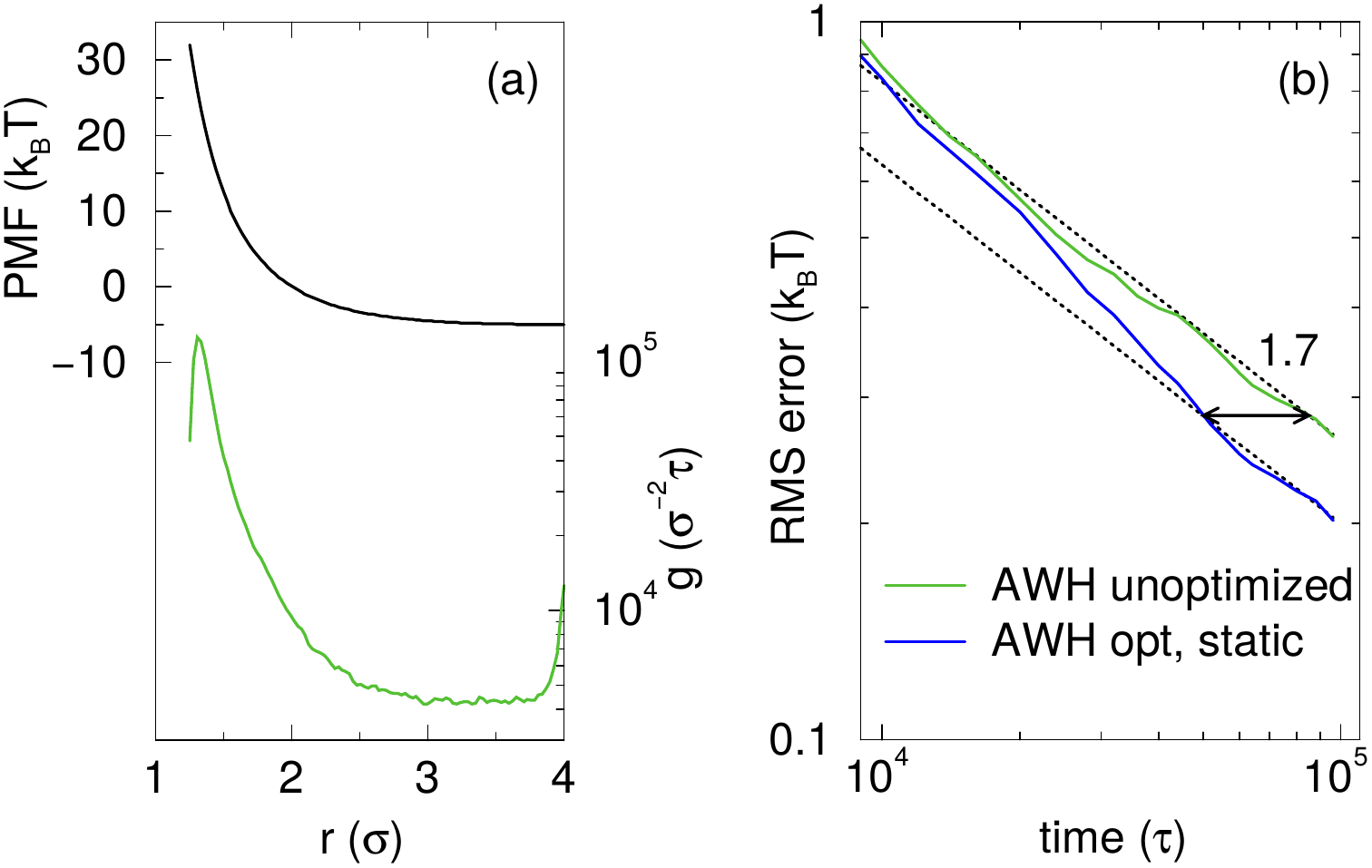}
\caption{PMF and metric for polymer-wall distance (a) and
convergence of the PMF (b).
The metrics for the unoptimized and optimized cases are identical.
Note that the peak in the metric at $r=4\sigma$ is an edge effect,
which decreases when increasing the force constant used in AWH.
} 
\label{fig:polymer}
\end{figure}

As a first example we present attaching/detaching a polymer chain to/from
a surface. The polymer is a freely jointed chain of Lennard-Jones (LJ)
beads and the reaction coordinate is the distance of the center of mass of
the polymer chain to the wall.
We will express the parameters in LJ units of length $\sigma$,
energy $\epsilon$ and time $\tau$.
The chain has 80 beads and a joint length of $\sigma$.
The surface is a 10-4 type potential,
$U(r) = (\epsilon \pi/4) (\sigma/r)^{10} - (\epsilon \pi/10) (\sigma/r)^{4}$,
obtained by integrating a LJ potential
over a plane with a surface density of 0.5$\sigma^{-2}$. The temperature
was set to 3~$\epsilon/k_B$ and $k_\mu$ to 1333$\sigma^{-2}$.
The reaction coordinate range was chosen as 1.25$\sigma$ to 4$\sigma$,
which goes from the shortest possible distance to a mostly detached polymer.
The system was simulated using Langevin dynamics with a friction
coefficient of 0.01$\tau^{-1}$ and an integration timestep of 0.001$\tau$.
The free energy varies over the large range of 47 (in units of $k_B T$); see Fig.~\ref{fig:polymer}.
The volume of the chain measured as the radius of gyration stays approximately
constant at 6 $\sigma$ in the whole sampling range. But to keep this volume,
the area parallel to the surface needs to increase inversely proportional
with the distance from the surface, which requires large conformational
rearrangements.
Indeed the metric shows a high peak at short distance and flattens out at larger
distance as the chain relaxes to its ``solution'' state. Thus to optimize
sampling along the reaction coordinate either a nonuniform target distribution or
a nonlinear transformation of the reaction coordinate is required.
As an error measure we used the free-energy difference between the points
neighboring the end points, 1.277$\sigma$ and 3.973$\sigma$, to avoid
the slightly more noisy endpoints which lack neighbors on one side.
The error, shown in Fig.~\ref{fig:polymer}, was estimated from the square root of the variance over 360 independent simulations.
The unoptimized case reaches an error of 1 at a time of $10^4\tau$
and shows the expected $1/\sqrt{t}$ convergence after that.
The statically optimized case, with the metric taken from the unoptimized case,
continues longer with faster convergence
and shows an efficiency improvement of a factor of 1.7.
This is higher than the factor 1.4 calculated from Eq.~\eqref{eq:improvement};
the difference is likely due to the locality assumption, which is violated
due to the slow and global nature of the conformational changes involved.

When using the dynamic doubling optimization protocol (described in Section~\ref{sec:numeric})
for this system we found only a small efficiency gain.
Although the metric estimate converges much faster than the free energy, a substantial fraction of the simulation time was spent on building up the estimate of the metric.
On the other hand, the statically optimized case reflects the asymptotic efficiency improvement for long simulations using dynamic protocols, where only a small fraction of the simulation time is used to build up the metric estimate.
Thus, for much longer simulations the dynamic and static cases are bound to approach each other,
and the static improvement factor represents the best convergence one could obtain with such an approach.
These observations apply also to the other test cases discussed below.
In addition, in practice the metric can often be estimated from pre-production data or from
prior simulations of similar systems, in which case static optimization would exploit that knowledge.

\subsection{Example II: Lithium acetate in water}\label{sec:liac}
\begin{figure}
\includegraphics[width=8cm]{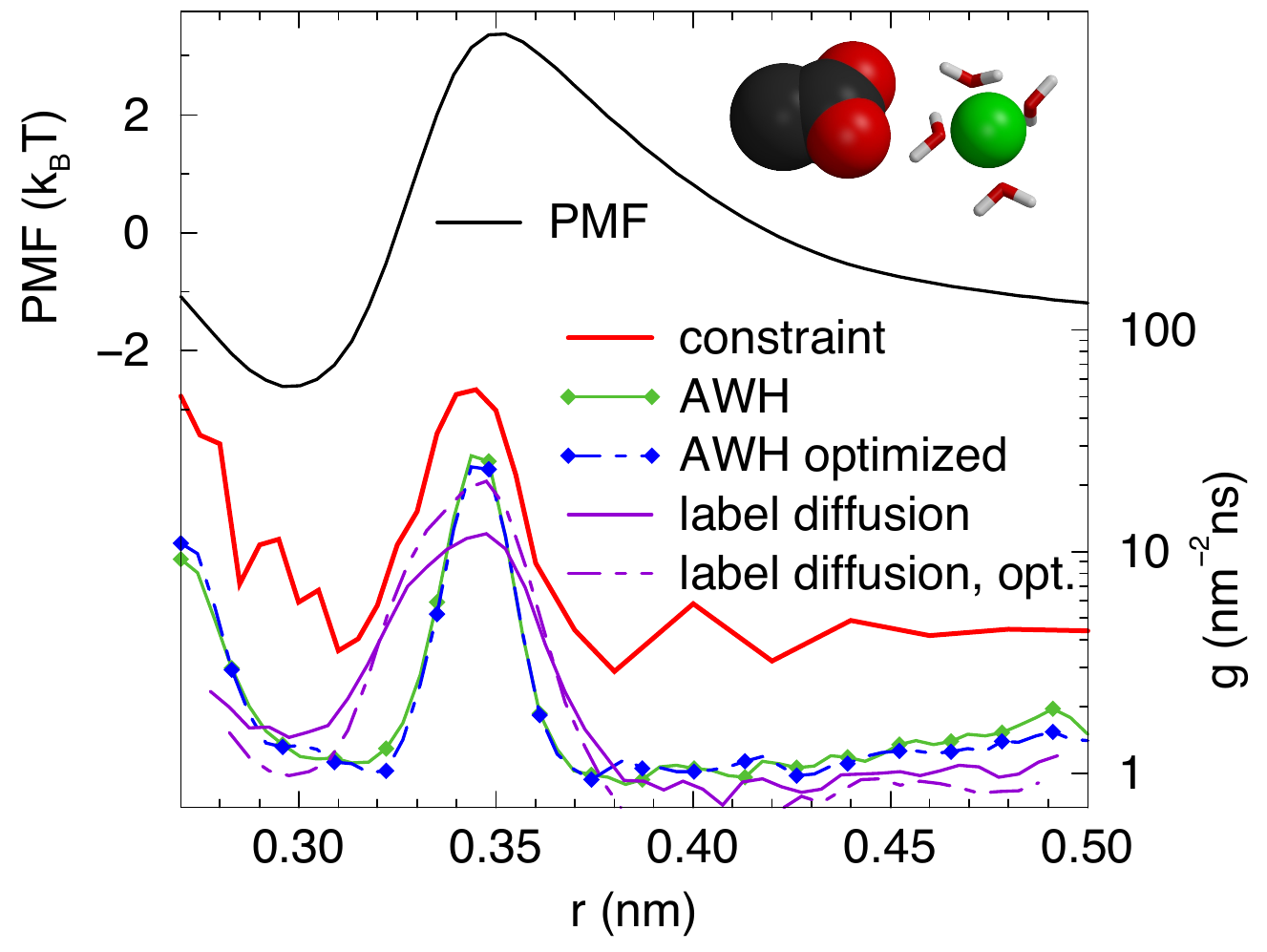}
\caption{The PMF and the metric along the distance ($r$) between 
a lithium and an acetate ion.
The solvent separated configuration at $r=0.5$~nm is also shown.
Note that the trivial entropic term $-2 \log(r)$ has been subtracted from the PMF.
The metric was calculated for two different sampling methods, AWH or constraints;
for AWH both using an optimized and unoptimized  target distribution. 
The metric $g=1/D$ was also calculated for an unoptimized and a self-consistently optimized AWH simulation
using the labeled walker approach to obtain the local diffusion constant $D$.
}
\label{fig:liac}
\end{figure}

As a second application we present a PMF calculation for separating a lithium 
and an acetate ion solvated in water. The reaction coordinate is the distance
between the lithium and the central carbon atom of the acetate,
see Fig.~\ref{fig:liac}.
 The model parameters and
the system setup were taken from work on optimized ion interactions~\cite{Hess2009}.
The range of the reaction coordinate
is from the contact pair (0.27 nm) to the solvent bridged pair distance
(0.5 nm; configuration shown in figure).  The force constant $k$ was set to 51,200 nm$^{-2}$.
To move between these states,
a free-energy barrier needs to be overcome that involves moving the bridging
water molecule in or out between the ion pair.

We determined the metric and the PMF using either AWH or the method of constraints combined with thermodynamic integration~\cite{straatsma1986free}.
As noted previously, for the latter our metric is equivalent to $g^{\text{SC}}_{\mu\nu}(\lambda)$, Eq.~\eqref{eq:Sivak-Crooks-metric}.
In the case of AWH, simulations were performed using either a non-optimized, $\pi(\lambda) = \text{const.}$, 
or  optimized, $\pi(\lambda) \propto \sqrt{\hat{g}(\lambda)}$,
target distribution.
We furthermore test the three different optimization  protocols described in Section~\ref{sec:numeric}:
\textit{static};
\textit{dynamic, continuous};
or \textit{dynamic, doubling}.
In addition, for comparison, we optimized AWH simulations
using the method of Trebst \emph{et al}, calculating $D(\lambda)$ using labeled walkers in an unoptimized AWH simulation as well a for a self-consistently optimized AWH simulation.
The consistent ``metric'' in this case is simply $1/D(\lambda)$.
The constraint runs had 0.1 ns equilibration and 4 ns data collection per point. 
For AWH, we generated  120 to 200 independent runs of 10 ns each per setup.

The PMF (obtained using the method of constraints) together with the different metric profiles
are shown in Fig.~\ref{fig:liac}. 
The most obvious, common characteristic of either metric is a peak close to the maximum of the PMF barrier.
At this distance the hydrogen bonding network around the two
ions needs to rearrange to accommodate for the bridging water moving in or out.
This involves movements of degrees of freedom orthogonal to the reaction coordinate, which results in longer correlation times and therefore higher metric.
The metric with the constraint method is approximately a factor 4 higher 
than with AWH. 
This can be explained by the fact that constraining a degree of freedom hinders
transitions along other degrees of freedom that could occur more frequently
by (slight) changes in the reaction coordinate. In contrast, the AWH method
allows free diffusion along the reaction coordinate.

To evaluate the effects on the metric of the magnitude of the generalized force
and the correlation time, we computed full autocorrelation forces for
unoptimized and optimized AWH runs. We observe approximately exponentially
decaying autocorrelation functions with a correlation time of 1 to 2 ps
at the base level and 25 ps at the peak in the metric, whereas the metric differs
by a factor of 20.
Thus, here the correlation time contributes more than the magnitude of the force fluctuations to the difference in metric.

\begin{figure}
%\centerline{\epsfig{file=plot/liac_fediff.eps,height=8cm}}
\centerline{\includegraphics[height=8cm]{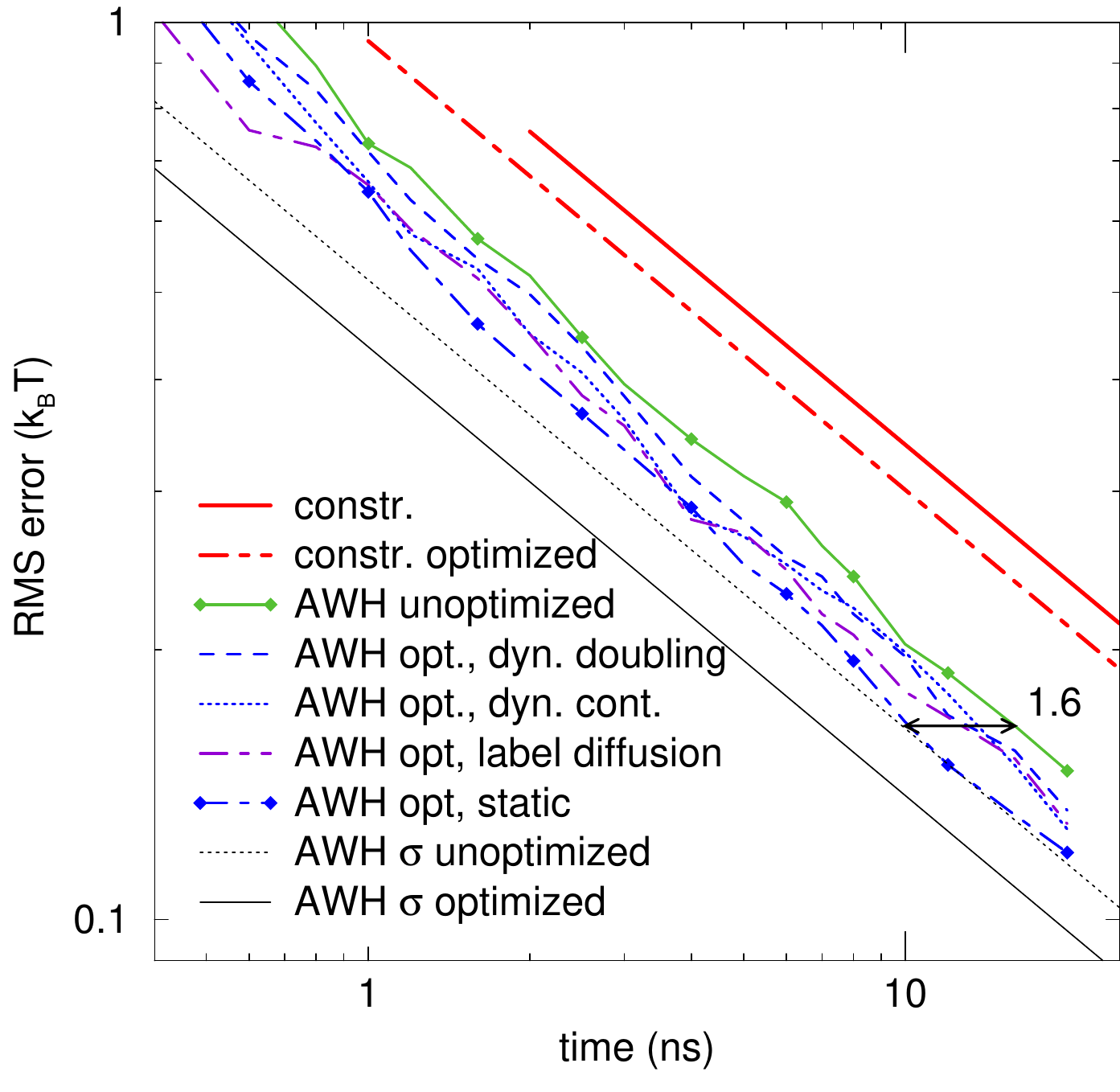}}
\caption{Convergence of the lithium acetate PMF difference between $r$=0.27
and $r$=0.50 nm
for different sampling methods (AWH or constraints) and optimization protocols.
The optimized target distribution $\pi(\lambda) \propto \sqrt{g(\lambda)}$,
where the metrics $g(\lambda)$ are shown in Fig.~\ref{fig:liac}.
The various optimization protocols are described in the main text.
For sampling using constraints
the average error,
for both optimized and non-optimized target distributions,
 is given as $\sigma$, see Eq.~\eqref{eq:sigma2}.
$\sigma$ is also shown for AWH sampling, for which it underestimates the actual error.
The largest improvement factor 1.6 (indicated by a horizontal double-headed arrow) 
is obtained for sampling with AWH using static optimization.
}
\label{liac_err}
\end{figure}

It is interesting to note that the change in target distribution, which also
affects the dynamics has negligible influence on our metric (see Fig.~\ref{fig:liac}). We have observed this for all systems we have studied.
This can be contrasted with the diffusion obtained from labeled walkers
where the height of the peak increases by a factor of two after optimization.
Thus a self-consistent optimization scheme is required to optimize for
this metric. For lithium acetate we found that it no longer changes after one iteration.
Statically optimizing AWH, with $\pi(\lambda)$ taken from unoptimized
simulations at 8 ns, improves the efficiency by a factor 1.6; see Fig.~\ref{liac_err}.
This improvement is slightly better than expected from the simple estimate 
${\sigma^2_{\mathrm{opt}}}/{\sigma^2_{0}}$,
see Eq.~\eqref{eq:improvement}.
Furthermore, we note the real errors are higher than $\sigma$
(also shown in the figure)
likely because the reduced dynamics is not fully Markovian.
The choice of dynamic optimization protocol also has a small effect on the convergence.
For this system the continuous scheme is stable and for most times the error is lower than for the doubling procedure, as one would expect.
Compared to optimized sampling using the method of constraints, 
statically or dynamically optimized AWH sampling  reduces the variance by a factor of 3.
This shows that choosing a sampling method with a dynamic reaction coordinate can improve sampling
significantly.

Improvement in sampling arises from faster exchange between different important states
of the system, which should correlate with the mean round trip time 
$\tau_{\mathrm{rt}}=\tau_{\mathrm{up}} + \tau_{\mathrm{down}}$,
where $\tau_{\mathrm{up}}$ and $\tau_{\mathrm{down}}$ 
are the mean first passage times going in the upward and downward direction, respectively~\cite{nadler2007generalized}.
On the contrary, Ref.~\cite{nguyen2016intermediate} did not find any significant correlation between the mean first passage time and accuracy in replica exchange MD simulations, possibly due to the restriction to nearest neighbor exchange.
In our case
 we see that indeed the round trip time
 decreases with the variance of the calculated free energies,
see Table~\ref{tabmfp}. 
Furthermore, optimization tends to equalize up and down times.
Under the  assumptions of the diffusion equation, this is a direct consequence of maximizing the flow~\cite{nadler2007generalized, abreu2009}.
For the unoptimized target distribution, 
 $\tau_{\mathrm{down}}$
is nearly
twice as long as
 $\tau_{\mathrm{up}}$.
This is likely mainly caused by the fact
that the upper boundary is further away from the region with high metric.
Optimizing the sampling using our metric increases 
 $\tau_{\mathrm{up}}$  slightly,
but lowers
 $\tau_{\mathrm{down}}$
much more.

Self-consistently optimizing using the diffusion from labeled walkers gives an equally low $\tau_{\mathrm{up}}$
but results in a larger $\tau_{\mathrm{down}}$.
This is consistent with the slightly larger error we obtain for this case compared to using our metric,
see Fig.~\ref{liac_err}.
We see from Fig.~\ref{fig:liac} that this difference must arise either from the wider peak at $r=0.35$~nm
or from the lower end $r<0.35$~nm.
We therefore also optimized with a target distribution equal to the 
labeled walker target distribution for $r> 0.29$~nm
but equal to the metric optimized target distribution further down.
As expected, this increases  $\tau_{\mathrm{up}}$
and reduces  $\tau_{\mathrm{down}}$  but leaves the round trip time  and the error unaffected.
Thus we conclude that the main difference originates from the peak region.

\begin{table}
\begin{tabular}{l|rrr}
Target distr. & Variance (k$_B^2$T$^2$) & $\tau_\mathrm{up}$ (ps) & $\tau_\mathrm{down}$ (ps)\\
\hline
Unoptimized        & 0.027 $\pm$0.003 & 66.4 $\pm$0.4 & 118.3 $\pm$1.0 \\
Opt, static        & 0.017 $\pm$0.002 & 74.9 $\pm$0.2 & 74.3 $\pm$0.2 \\
Opt, label diff.   & 0.023 $\pm$0.002 & 74.3 $\pm$0.4 & 82.6 $\pm$0.5 \\
\end{tabular}
\caption{Mean square error of the PMF and mean first passage times $\tau$ for the lithium acetate ion system sampled using AWH and three different target distributions.
The mean square error was calculated for the PMF difference between the end points of the sampling interval at 16 ns.
The mean first passage times in both directions, $\tau_{\text{up}}$  and $\tau_{\text{down}}$, are brought closer together by using the optimized target.
\label{tabmfp}}
\end{table}

\subsection{Example III: DNA base pair opening}\label{sec:dna}
\begin{figure*}
\includegraphics[height=4.8cm]{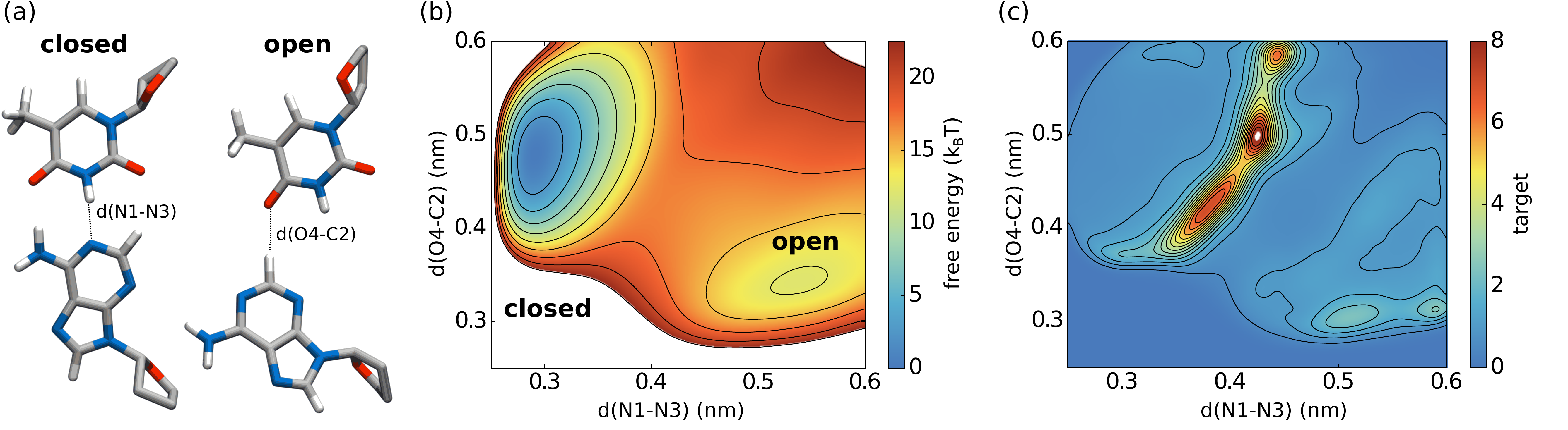}
\caption{
DNA base pair opening.
The closed, WC state is characterized by low values of  $d(\text{N1--N3})$ (a), 
the distance between the atoms that form the middle WC hydrogen bond.
Upon base pair opening, $d(\text{N1--N3})$ increases, 
which favors a non-WC interaction,
characterized by a short distance $d(\text{O4--C2})$.
The PMF landscape (b) has two minima,
the global minimum corresponding to the closed state
and a second local minimum, corresponding to open conformations.
The white region was excluded from sampling by using a free-energy cutoff.
The optimized target distribution (c), given by Eq.~\eqref{eq:target},
is sharply peaked in a transition region between the two minima.
}
\label{dna_fe}
\end{figure*}

As a third, more challenging application we present DNA base pair opening.
The most common state of DNA is the double helix where every base pair 
interacts through Watson-Crick (WC) hydrogen bonds. But for its function, be it
DNA replication, modification or repair, 
the base pairs need to open, allowing the bases to flip out~\cite{frank2014fluctuations}.
He were study the initiation of base flipping 
for a periodically connected sequence of TCTAT\textbf{T}TATT and its complement,
where we open the sixth base pair (shown in bold type). 
The Amber parmbsc1 force-field \cite{ivani2016parmbsc1}
was used.
As a reaction coordinate for the opening, 
we used the distance of
the middle WC hydrogen bond donor and acceptor nitrogens in a T-A pair,
$d\text{(N1--N3)}$, see Fig.~\ref{dna_fe}.
We previously observed~\cite{lindahl2017sequence} that during the opening a new favorable interaction forms
between the O4 oxygen and the C2 carbon that complicates the sampling, 
%(see Fig. \ref{dna_fe}) 
since it is not
well aligned with the WC hydrogen bond reaction coordinate. Therefore
we added  the distance
$d(\text{O4--C2})$
 as a second reaction coordinate.
Both coordinates are sampled from 0.25 to 0.60 nm with a harmonic force
constant of 51,200 nm$^{-2}$.
To avoid regions of high free energy, which can lead to unphysical states,
we use target distribution Eq.~\eqref{eq:target-cutoff} with the sigmoidal cutoff function $\varphi(z)=1/(1+\exp(z))$, where the free-energy cutoff 
$F_{th} = \min_\lambda F(\lambda) + 20$.
We ran 120 unoptimized AWH simulations of 160 ns each. 
The average metric 
at $t=100$~ns of these unoptimized runs
was used to statically optimize 120 simulations, each 120 ns long.
We also ran 200 dynamically optimized simulations, 120 ns long 
using the previously defined doubling optimization protocol.
It turns out that this two-dimensional AWH sampling converges faster than the 
one-dimensional case where only $d\text{(N1--N3)}$ is biased.

Because of the complex base pair opening mechanism, and the likely suboptimal
reaction coordinate, a small fraction of the simulations shows poor convergence.
To avoid these problematic runs from dominating the results, we excluded
runs where the empirical distribution differed on average by more than a fraction 0.55
from the target distribution at an average error of $\approx$\,0.85
(corresponding to times 140, 120 and 100~ns for the unoptimized and dynamically and statically
optimized runs, respectively).
Note that this criterion does not directly involve
the (converged) free energy or the error. This excludes 4\% of the simulations.
Using a tighter criterion excludes more simulations but does not change
the average error.
Such a check is useful in general for applications of
histogram-based adaptive methods.

\begin{figure}
\includegraphics[height=6cm]{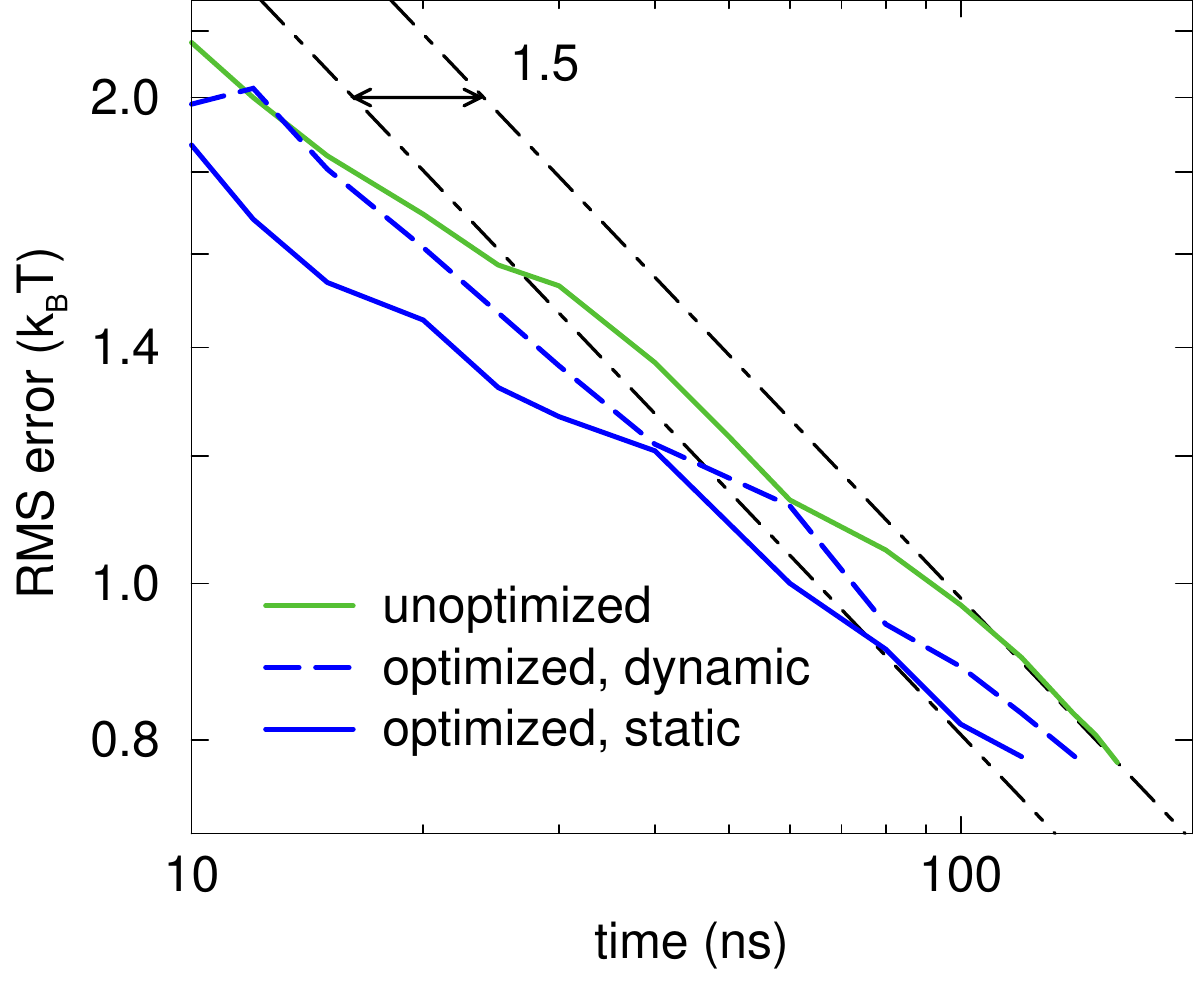}
\caption{The convergence of the free energy for DNA base pair opening for different optimization protocols.
AWH was used to calculate the metric yielding the optimized target distribution,
as well as the two-dimensional PMF, see Fig.~\ref{dna_fe}.
The error is based on the free-energy difference between the two local minima of the PMF.
The simulation length was set to give a final error of $\approx$\,0.8.
The maximum improvement factor of 1.5 (indicated by a horizontal double-headed arrow) was obtained using static optimization.
The straight lines indicate the long-time convergence rates for the unoptimized and statically optimized cases.
}
\label{dna_err}
\end{figure}

The free-energy landscape and the target distribution are shown in Fig. \ref{dna_fe}.
We observe two, hydrogen bonded, minima that differ in free energy by 12.0.
A high peak in the metric separates the two minima.
Like in the case of lithium acetate, most of the friction arises from
rearrangement of the hydrogen bonding network.
We computed the error in the free-energy difference between
the global WC minimum and the second, local minimum.
The convergence is shown in  Fig. \ref{dna_err} for
the unoptimized and optimized target distribution using the doubling interval optimization protocol.
In this case, optimizing the target distribution reduces the required simulation time by up to a factor of 1.5.

\section{Conclusions}

Extended or generalized ensembles, where one or more system parameters are promoted to dynamical variables, or where suitable reaction coordinates are used to guide the system through a transition,
are highly useful for enhancing sampling of systems with complex energy landscapes. 
The dynamics of the original very high dimensional system is thereby projected onto a much reduced space, with in general a non-Euclidean geometry.
We have introduced a suitable Riemann metric to describe the geometric properties of this parameter manifold $\Lambda$, and have shown that the choice of parameters and their marginal target distribution may be guided and optimized by these geometric considerations.
For instance, the geodesics form optimal pathways for evaluating free-energy differences between two states.
Further, by defining the target distribution in terms of the metric it becomes reparametrization invariant.
Without a proper metric the target distribution will instead depend on the parametrization in an arbitrary way.

In a one-dimensional setting, the variance of an estimated  free-energy difference between two points in $\Lambda$ is minimized by distributing the samples uniformly over the arc length.
In higher dimensions we propose to use a uniform target distribution \emph{with respect to the metric}, Eq.~\eqref{eq:target}, to allow the system to freely explore multidimensional regions.
This comes at a price, however, since some importance sampling is quickly lost when sampling uniformly in high dimensions.
There is also the risk that the metric may amplify the target distribution in uninteresting regions, in case they are hard to sample.
Some ideas for further restricting sampling to interesting regions are contained in Eqs.~\eqref{eq:target-cutoff}-\eqref{eq:target-distance-2}, by introducing a free-energy cutoff or by sampling within a metric distance from some suitable point(s).

The metric itself is a locally defined quantity that can be estimated reliably without requiring extensive global sampling.
This stands in contrast to the diffusion optimized labeled walker approach of Trebst \textit{et al.}~\cite{trebst2004optimizing}, which moreover is limited to one dimension.
In the lithium acetate test case, studied in detail above, we found a slightly sharper target distribution and somewhat better accuracy when optimizing using our metric.
More important is that convergence of our metric depends only locally on the amount of sampling,
which makes adaptive updating of the target distribution easier and more robust.

In the present work, we have focused on the application to reaction coordinates in MD simulations. 
Using the AWH method to adaptively apply a bias potential,
we have demonstrated how to carry out the optimization in a fully automated fashion,
at negligible extra computational cost.
In the three examples we have presented, a polymer at a wall, lithium acetate in water, and DNA base pair opening, we found an increase of sampling efficiency of 50--70\%
for the static optimization case, where an estimate of the metric was assumed to be known from the start of the simulation. These numbers reflect the asymptotic improvement achievable for long simulations.
If the metric is instead calculated and applied on the fly, the improvement factor may be smaller since part of the simulation will be spent on building up the metric estimate.
When prior knowledge is available or when many similar simulations are generated an advantageous approach  is therefore to apply a static target distribution.
For very long simulations the choice of optimization protocol will be less important.

The amount of speedup gained by optimizing the target distribution depends strongly on how much the metric varies over the parameter region, see e.g.\ Eq.~\eqref{eq:improvement}, and is thus highly problem dependent.
The cases studied here all converged well also using an unoptimized, i.e. uniform,  target distribution, allowing us to make these comparisons. More difficult cases, e.g. sampling of phase transitions~\cite{trebst2006optimized}, will have a greater potential for speedup.

We have focused here on sampling along reaction coordinates.
The metric may also be used to optimize, e.g.,
ensembles extended along temperature or energy, or for alchemical transformations.
We described specifically how to optimize the target distribution in AWH simulations, but
other enhanced sampling methods could be used instead, e.g.,
Wang-Landau~\cite{wang2001efficient} or metadynamics~\cite{laio2002escaping},
in particular its variational formulation~\cite{valsson2014variational}, where the target distribution can be prescribed.
Similar geometric considerations are relevant also for thermodynamic integration and for replica exchange (parallel tempering) simulations.

The choice of parameters or reaction coordinates $\lambda$ remains the most difficult and critical step in applications of enhanced sampling.
The metric
can in this respect be helpful in identifying bottlenecks and other difficult-to-sample regions in parameter space.
A peak in the metric is often an indication that degrees of freedom orthogonal to the reaction coordinate(s) may be important.
Choosing a target distribution as in Eq.~\eqref{eq:target} automatically allocates samples to compensate for such misalignment issues.
However, should extreme variations in the metric arise, one might consider other choices for the definition of $\lambda$,
which potentially avoid such bottlenecks.

To conclude, we have presented a Riemann metric on the multidimensional space of parameters or reaction coordinates,
that takes time correlations into account,
and which provides a practical and general way to help decide how the samples should be distributed among the parameter values.
Furthermore, the metric opens up new possibilities to guide sampling in multidimensional free-energy landscapes.

\begin{acknowledgments}

This research was supported by the European Research Council (grant
no.~258980) and the Swedish Research Council (grant no.~2014-4505).
The simulations were performed on resources provided by the Swedish
National Infrastructure for Computing (SNIC 2016/1-562, SNIC 2016/10-47 and 2017/11-25) at
the PDC Centre for High Performance Computing (PDC-HPC)
and the High Performance Computing Center North (HPC2N).

\end{acknowledgments}

% References
\bibliography{art}{}
\end{document}